# The State of Scientific Poster Sharing and Reuse

## Authors

Aydan Gasimova[+,1], Paapa Mensah-Kane[+,2], Gerard F. Blake[3], Sanjay Soundarajan[1], James O'Neill[1], Bhavesh Patel[1,*]

[+]Have contributed equally and are listed alphabetically by last name

## Affiliations

[1]FAIR Data Innovations Hub, California Medical Innovations Institute, San Diego, CA, USA
[2]Department of Pharmaceutical Sciences, School of Pharmacy, South University, Savannah, GA, USA
[3]The University of Chicago, Chicago, IL, USA

***Correspondence author**:
Bhavesh Patel
Research Professor
FAIR Data Innovations Hub
California Medical Innovations Institute
11107 Roselle St.
San Diego, CA 92121
T: (510) 604-2815, Email: bpatel@calmi2.org

## Abstract
Scientific posters are one of the most common forms of scholarly communication and contain early-stage insights with potential to accelerate scientific discovery. We investigated where posters are shared, to what extent their sharing aligns with the FAIR principles, and how commonly they are reused. We identified 86 platforms hosting posters, with many not assigning persistent identifiers. A total of 150k posters are shared as of 2024 on the 43 platforms where we were able to count, which is relatively low. Looking in more detail at posters shared on Zenodo and Figshare, we found that repositories are not always supporting structured metadata critical for poster discovery, like conference information, and that researchers are not providing such metadata even if they are supported. We also observed that while there is some engagement with posters in terms of views and downloads, citing posters is not yet a common practice. These gaps mean valuable early-stage insights are being lost to the scientific community, pointing to a need for clearer guidelines and stronger incentives for FAIR poster sharing and reuse.

## Introduction

Every year, scientific conferences bring together millions of researchers to showcase discoveries, exchange ideas, and build collaborations.[1] Amongst the many forms of communication at these events, posters are especially prevalent.[2] While no exact count is available, it is estimated that millions of scientific posters are presented every year, making them a major form of scientific communication.[1] Posters often highlight early-stage insights and preliminary results, introducing them for the first time beyond the research team.[1] As such, they represent a valuable source of scientific knowledge.

Despite their prevalence, only a fraction of posters are ever developed into peer-reviewed manuscripts. Across different disciplines and meetings, reported publication rates for posters typically vary widely from as low as 2% to around 50%, with many studies consistently showing that posters are less likely than oral presentations to progress to full publication.[3,4,5,6,7,8] Even when they do, there is often a delay of several years between poster presentation and manuscript publication.[9] This means that a substantial amount of knowledge shared at conferences never becomes part of the permanent scientific record or takes a long time to do so, potentially limiting its visibility and long-term impact. Expanding opportunities for sharing posters beyond the conference setting could help preserve their value and extend their impact beyond the few days of a meeting and the limited audience in attendance.

Integrating posters into the broader digital research ecosystem, rather than treating them as isolated conference artifacts, can help address this gap. The FAIR (Findable, Accessible, Interoperable, Reusable) principles provide a widely adopted framework for ensuring that research outputs can be effectively discovered and reused, in particular through persistent identifiers and rich metadata. Applying these principles to posters is particularly important as it would extend their lifespan and visibility. Put plainly, the tools researchers now use to find work, from search engines to Artificial Intelligent (AI)-based assistants, can only surface a poster they can read and without metadata a shared poster stays effectively invisible to them. Making posters FAIR would also allow their content to be incorporated into digital knowledge infrastructure alongside other research outcomes, such as manuscripts and datasets. Additionally, this could also give authors of posters, who are often early-stage researchers, greater recognition and opportunities for collaboration. The emergence of general-purpose repositories such as Zenodo and Figshare has made poster sharing technically feasible. However, it is still unclear how common poster sharing and reuse are and to what extent poster sharing practices align with the FAIR principles.

To address this gap in knowledge, we investigated where scientific posters are shared digitally, how their sharing aligns with the FAIR principles, and how often they are reused. To achieve this, we conducted a broad online search to find platforms where posters are typically shared. Since, to our knowledge, there are no established guidelines for making posters FAIR, we focused on evaluating major requirements of the FAIR principles, such as having a globally unique and persistent identifier and having rich metadata to enhance findability. The goal of this work was to provide an evidence-based assessment of current practices so stakeholders, including researchers, conference organizers, repositories, and funders, can work towards improving poster sharing and reuse practices to maximize the long-term impact of these valuable research outputs.

# Methods

## Poster sharing

### Finding poster-sharing platforms

To our knowledge, no official list or registry of platforms where posters are shared exists. We therefore used three methods to identify platforms where posters are shared.

Method 1: Google Search

We began by searching through Google, aiming to identify guidelines from conferences, universities, or institutions that recommend platforms for sharing posters. The search was conducted on March 4th, 2025. A Python script was developed using the googlesearch-python package and executed in a Jupyter Notebook to automate the retrieval of search results.[10,11] We used “poster sharing publication” as the search query and manually screened the results to identify any platform that was mentioned or suggested for sharing scientific posters. We noticed that after the 91st result, the results were not relevant, so we did not continue after the first 100th result. The code used for this search is provided in the “find-posters.ipynb” notebook, which is included in the code associated with this work (see the **Code Availability** section). The search results, a list of all platforms found, and additional details about the search are included in the “poster-platforms-search-Google.xlsx” file included with the data associated with this work (see **Data Availability** section).

Method 2: Large Language Models

We included Large Language Model (LLM) queries as an additional strategy to see if they could suggest additional platforms for sharing posters that may not be easily discoverable through a conventional search engine like Google. We queried two Large Language Models (LLMs): OpenAI's ChatGPT (version 4o mini) and Google's Gemini AI (version 1.5 Flash).[12,13] The queries were conducted on March 4, 2025, on the respective web platforms of these models. We followed similar prompting strategies for both, where we started with an initial prompt (“Give me a list of all the platforms where I can share a scientific conference poster in a way that others can access it”) and followed up with additional prompts depending on the initial answer to extract as many suggestions of potential poster sharing platforms as possible. The prompt, responses, a list of platforms found, and additional details about the search are included in the “poster-platforms-search-LLM.xlsx” file (see **Data Availability** section).

Method 3: DataCite

Many of the platforms identified through Methods 1 and 2 assign a Digital Object Identifier (DOI) to their posters through the DataCite platform.[14] Therefore, we searched for posters in the DataCite database using their API.[15] While there is no official entry in the DataCite schema to register a resource as a poster at the time of our search, we noticed that most of the platforms found through Methods 1 and 2 register posters through DataCite using “Poster” as a free text for the resource type field. To identify additional DOI issuing platforms that share posters, we used the DataCite API to search through all their records for resources registered as “Poster” and extracted the names of their publishers. The search was conducted on March 8th, 2025.

The related code is included in the “find-posters.ipynb” Jupyter notebook. The search results, a list of potential platforms found, and additional details about the search are included in the “poster-platforms-search-DataCite.xlsx” file included with the data associated with this work (see **Data Availability** section). Note that this approach is likely not identifying all the platforms registering a DOI for posters since some platforms may not set the resource type to “Poster”. For instance, the F1000Research platform registers DOIs for posters using “Other” as the resource type.[16]

**Evaluating the platforms**

We compiled a list of unique platforms identified through these three methods and reviewed them to verify if they contain scientific posters or not. Each platform was randomly assigned to a team member, and they randomly checked up to 10 resources classified as “poster” (if any) on the platform. We considered a platform to be a poster-sharing platform and retained it for our analysis if at least one of those resources was the file of a scientific poster (i.e., the file of the visual poster displayed at a conference). We excluded platforms hosting only abstracts or related papers but not poster files. We also excluded derivative platforms duplicating content from a parent repository (e.g., institutional portals based on Figshare) to prevent duplicate count in our analysis. For each platform retained, the assigned team member recorded the following information in a spreadsheet: 1) Link to the platform, 2) Link to a sample poster, 3) Whether it is exclusively for posters or also hosts other research outputs, and 4) Whether it assigns a globally unique and persistent identifier like a DOI to posters (by checking the sample poster). This is provided in the “poster-platforms-results.xlsx” file included with the data associated with this work (see **Data Availability** section).

**Counting posters**

To assess how commonly posters are shared, we also counted the total number of posters shared on each platform. We included in the count any poster whose latest version was shared on a platform on or before December 31st, 2024. The count was conducted by the assigned team member as follows, depending on what was possible for a given platform:

1. Where available, we used the platform’s search interface, applying filters to only keep posters and set the creation date (on or before December 31, 2024). On platforms without a formal date filter, we attempted several workarounds, such as editing the date directly in the search bar or using the query string.
2. For platforms without relevant filters, we manually counted posters when it was reasonable (fewer than 100 posters).
3. When none of these approaches were possible for a given platform, we did not count.

Note that for DOI-issuing platforms, we attempted to use the DataCite API to count posters, but the inconsistencies in linkage between different versions of a poster made it difficult to count each poster uniquely, so we excluded this approach.

During our search, we identified numerous institutional repositories built on the Digital Commons platform by Elsevier.[17] Upon looking further, we found that there are potentially over 600 such repositories.[18] Because they are operated independently through their respective websites, it would have been difficult to analyze each one separately. We therefore grouped all of them under a single entry labeled “All Digital Commons” in our list of poster-sharing platforms.

We examined a few of them manually, such as the repositories from Technological University Dublin, Utah State University, and Old Dominion University.[19–21]We found that they contain scientific poster files, but do not assign globally unique and persistent identifiers to them. On that basis, we made the assumption that other Digital Commons repositories follow similar practices and reflected this collectively in our analysis. The search interface of the repository from Technological University Dublin provides a feature to search across all Digital Commons repositories, so we used it to get the total poster count.

To examine trends in poster sharing over time, we also collected the number of posters shared on a platform over the years. This year-by-year count was conducted only for platforms where we were successful in obtaining the total number of posters shared, and only if that number was greater than 100 posters. For some platforms, it was not clear if the provided date was the date the poster was shared/uploaded on the platform or if it was the year the poster was presented. Such platforms were excluded from the year-by-year count. For each retained platform, the assigned team member conducted the year-by-year count directly from the platform's website by changing the date filter and looking at the total number of results for each year.

### Platform review and verification process

Each platform's results were verified by another team member (the second reviewer) independently. The verification process included: 1) Confirming that the platform link was active, 2) Checking the sample poster is a scientific poster file (and not just its abstract, for instance), 3) Checking if the platform provides a globally unique and persistent identifier by reviewing the sample poster, 4) Confirming whether the platform is a derivative or independent, and 5) Recounting posters (total and year-by-year, as applicable). If the second reviewer observed any discrepancies for a platform, they discussed them with the first reviewer to resolve them and update the record accordingly for that platform in the "poster-platforms-results.xlsx" file.

### Analysis of the results

We imported the "poster-platforms-results.xlsx" file into a Jupyter notebook to analyze our findings, such as the number of platforms found through each of the three methods, the number of posters each platform contains, the evolution of the number of posters shared over the year, and so on. The related code is available in the find-posters.ipynb notebook included with the code associated with this paper (see **Code Availability** section). Python libraries such as pandas, NumPy, Matplotlib, and seaborn were used to analyze and visualize the results.[22–25]

## Analysis of FAIR practices

Since, to our knowledge, there are no established guidelines for making posters FAIR, we assessed compliance with key FAIR elements, focusing on persistent identifiers and metadata practices.

### Persistent identifiers

During our review of poster-sharing platforms, we manually inspected at least one sample poster per platform to determine whether a globally unique and persistent identifier (e.g., DOI) was assigned. This evaluation is reported in the "poster-platforms-results.xlsx" spreadsheet. For

DOI-issuing platforms, we confirmed that the DOI of the sample poster resolved correctly to the poster landing page.

**Metadata collected by platforms**

Given their large collections of posters and the availability of open, well-documented APIs, our detailed metadata analysis was focused on the Zenodo and Figshare platforms.[26,27] We first looked at the structured metadata that these platforms collect from a user when they are sharing a poster. We defined structured metadata as metadata that can be entered into dedicated fields or forms (e.g., “Author name,” “Conference name”). This step allowed us to compare platform-level support for metadata elements, including which fields were mandatory versus optional. To achieve that, we simulated sharing a poster on each of these platforms and collected a list of metadata that is possible to provide, along with their requirements (mandatory vs optional). We documented this in the “crosswalk-zenodo-figshare.xlsx” file that is included in the dataset associated with this work (see **Data Availability** section).

**Completeness of metadata provided by researchers**

To evaluate what metadata researchers actually provide, we retrieved records of all available posters on Zenodo and Figshare as of December 31, 2024. A Python-based Jupyter notebook was developed to collect the metadata of each poster using the APIs of the two platforms.[28,29] This notebook, called “poster-metadata.ipynb”, is included in the code associated with this work (see the **Code Availability** section). We ran the code during January 2026. For Zenodo, we used the search endpoint with the “resource_type” set to “poster”. It returned the metadata record of 24,734 posters. For Figshare, we used the search endpoint with the “item_type” set to “poster”. This returned the partial metadata of 14,513 posters. To get the full metadata, we had to use the “articles” endpoint for each poster using their Figshare ids from the partial metadata. We found that the full record did not exist for 126 posters (for instance, because they were removed/deleted), and we were to obtain the full metadata for 14,387 posters. The metadata for each poster from Zenodo and Figshare (link, title, keywords, authors, etc., as available) was recorded in an NDJSON file called “zenodo.ndjson” and “figshare.ndjson”, respectively, for subsequent analysis (these files are available in the dataset associated with this work). Because Zenodo and Figshare use different internal labels for metadata fields, we created a harmonized schema by mapping repository-specific fields to standardized designations (e.g., “main_title” in Zenodo and “title” in Figshare were both mapped to “title”). To facilitate analysis, we created tables summarizing the metadata for each poster, with columns such as “has_license” (True/False), “authors_count”, “authors_affiliation_count”, and so on. These tables are included in the “zenodo.csv” and “figshare.csv” files available in the dataset shared with this paper (see **Data Availability** section). Python libraries such as pandas, NumPy, Matplotlib, and seaborn were used to analyze and visualize the results.[22–25] We particularly analyzed the completeness of what we called general metadata, i.e., metadata that are typically expected for FAIR digital objects like data: title, version number, license, author names, author affiliations, author ORCIDs, description, keywords, funding information, and references. We also analyzed the completeness of conference-related metadata supported by Zenodo: name, acronym, website, dates, location, and poster session information.

## Poster reuse

To evaluate poster reuse, we looked at three metrics for the posters shared on Zenodo and Figshare: number of views of the poster landing page, number of downloads, and number of citations. Our analysis was focused on open access posters with a DOI, which represented 38,147 posters in total, including 24,304 posters from Zenodo (98.3% of total posters on Zenodo as of December 1st, 2024) and 13,843 posters from Figshare (96.2% of total posters on Figshare as of December 1st, 2024).

### Views and downloads

The number of views and downloads were collected from Zenodo and Figshare during January 2026. These metrics were already included in the poster metadata harvest from Zenodo. For Figshare, we used the “views” and "downloads" endpoints to collect these metrics for each poster based on their Figshare ids (see the “poster-metadata.ipynb” notebook for the code). During the data harvesting process, we encountered a number of false zero values because reuse metrics are not exposed through their general API for posters hosted on institutional Figshare repositories. To address this issue, we used alternative URL formats provided by Figshare for institutional posters. However, retrieving usage statistics for some institutional repositories was challenging, since Figshare does not provide a publicly available list of all institutional URL identifiers. As a result, certain records could not be resolved automatically. For these cases, we implemented a manual web-scraping approach to obtain the missing usage statistics. Results were combined from the “zenodo.csv” and “figshare.csv” tables into a single “combined-metadata.csv” table, and view and download counts were added.

### Citations

We used OpenAlex and DataCite to find potential citations to the posters based on their DOIs.[14,30] For each citation, we saved the DOI of the poster being cited, the link of the citing source, and its publication year, which we used as a citation year, in a “poster-citation.ndjson” file (included in the dataset). The full code is available in the “poster-reuse.ipynb”, and “poster_reuse_analysis.py” files provided with the code shared with this work (see **Code Availability** section). For each poster, we first queried the OpenAlex API to find citations, and saved the poster DOI along with citation link, and citation year by OpenAlex. We then queried the DataCite API to find citations. DataCite only provides the link of the citing sources. If a link was not already in the list of citations identified through OpenAlex, we queried the OpenAlex, Crossref, and DataCite APIs in that order of priority (whichever succeeded first) to find the publication year for that citing source.[31] If none were successful, we saved just the DOI of the poster being cited and the link of the citing source without the citation year. The total number of citations per poster was then added to the “combined-metadata.csv” table.

A temporal validation procedure was applied to exclude false-positive citations, defined as instances in which the publication year of the citing work preceded the poster’s creation year. For each citation, we retained the data source (OpenAlex or DataCite), citation link, publication year, and citation type. The updated poster-level dataset was saved as “combined-table-analyzed”.csv.

### Analysis

We used the SciPy Python library to explore the distribution of views, downloads, and citations.[32] We characterized the distribution using descriptive statistics, including central tendency (mean, median), dispersion (standard deviation (SD), percentiles P25 through P99), and extremity (min, max). Given the high volume of posters with zero values, particularly for citations, we also calculated the sparsity of the poster reuse metrics (number of posters with zero values, distribution of top shares to quantify the degree of concentration). Descriptive statistics were computed for all engagement metrics, including measures of central tendency, dispersion, and concentration (top 1%, 5%, and 10% share of total engagement).

We also analyzed whether richer metadata was associated with higher poster views, downloads, and citations. The metadata variables from the "combined-metadata.csv" table retained for the analysis are provided in **Table 5**. Given the different nature of their distributions, we opted for different analytical frameworks for views and downloads on one side and citations on the other. For both, we first conducted bivariate analysis to understand the impact of individual metadata variables, followed by multivariate modeling to evaluate the combined influence and interactions of these variables.

Views and downloads

While the vast majority of posters have at least one view or download, their distributions are heavily right-skewed and characterized by overdispersion. Given this non-normal distribution and the presence of extreme outliers, our correlation analysis was conducted as follows:

- Bivariate analysis for continuous metadata variables: To assess the relation between views/downloads and continuous variables (like percentage of authors with affiliations, number of keywords, number of words in the description), we used Spearman's Rank Correlation. Converting raw view/download counts into rank allowed us to identify monotonic relations between a variable and views/download without letting outliers in the top 1% of posters heavily influence the results.
- Bivariate analysis for boolean metadata variables: For boolean variables (for instance, "has_license", "has_funding_information"), we applied the Mann-Whitney U Test. This test compares the rank sum of two groups (e.g., posters with funding information and posters without) to determine if one group consistently ranks higher than the other without assuming a normal distribution of the underlying counts.
- Multivariate analysis (feature interaction): To evaluate the simultaneous influence of multiple metadata features, and account for potential interaction, we used Negative Binomial Regression with the dispersion parameter ($\alpha$) estimated from an intercept-only model using the method of moments ($\alpha$ = (variance - mean) / mean$^2$) and held fixed during coefficient estimation. This approach was selected because the extreme overdispersion in our data (variance/mean > 5,000) caused convergence failures when attempting to estimate $\alpha$ jointly with coefficients. Fixing $\alpha$ ensures model convergence while maintaining consistent coefficient estimates.[33] Continuous predictors were standardized (z-scored) prior to model fitting to ensure numerical stability. Results are reported as Incidence Rate Ratios (IRR) with 95% confidence intervals, where IRR represents the multiplicative change in expected views or downloads for a one-unit (or one SD) increase in the predictor.

Citations

The distribution of citations is characterized by severe zero-inflation and extreme concentration. Therefore, we followed a two-part analysis to distinguish between metadata variables that are important for getting cited and those that drive higher citation counts amongst posters that are cited:

- Bivariate analysis of citation probability: To identify which individual metadata variable correlates with a poster being cited at all, we converted citation counts into binary outcomes (0: not cited, 1: cited). For continuous variables, we used the Point-Biserial Correlation and for boolean variables, we used the Chi-Square Test of Independence. These tests measured the metadata's association with the likelihood of receiving any citation.
- Bivariate analysis of citation: For the subset of posters with at least one citation (n = 1,955), we conducted Spearman's Rank Correlation for continuous variables and the Mann-Whitney U Test for boolean variables. This allowed us to analyze which metadata variable influences higher citation independently of the non-cited posters.
- Multivariate analysis (feature interaction): To evaluate the combined effect of metadata variables, we employed a two-part hurdle model rather than Zero-Inflated Negative Binomial (ZINB) regression. This decision was made because the extreme overdispersion among cited posters (variance/mean ratio exceeding 1,800) caused ZINB convergence failures. The two-part hurdle model consists of: (1) a logistic regression predicting the probability of receiving any citation (Part 1), and (2) a Negative Binomial regression with fixed dispersion parameter predicting citation count among the subset of cited posters (Part 2). This approach is more interpretable than ZINB as it explicitly models citation probability and citation accumulation as separate processes with potentially different predictors. Part 1 results are reported as Odds Ratios (OR) and Part 2 results as Incidence Rate Ratios (IRR), both with 95% confidence intervals.

## Results

### Poster sharing

We identified 151 unique platforms where scientific posters are potentially shared. Upon reviewing them, we found 47 of them did not contain scientific posters, and 18 of them were derivative, duplicating content from a parent repository (e.g., institutional portals based on Figshare or Digital Commons). After reviewing the remaining ones, we identified 86 unique platforms that contain scientific posters and were retained for subsequent analysis (**Fig. 1**). DataCite search returned the largest number of relevant platforms (66), followed by LLMs (9), then Google (5).

We were able to reliably get the total number of posters on 43 platforms. They contain 150,002 posters as of December 31st, 2024 (our cut-off date for the analysis). Digital Commons, Zenodo, Figshare, and F1000Research host the largest number of posters, accounting for 89.8% of all the posters counted (**Fig. 2a**). We were able to reliably get yearwise counts for 4 platforms: Zenodo, Figshare, F1000Research, and NASA Technical Reports Server (NTRS). We observed that the number of posters shared is globally increasing over the years (**Fig. 2b**).

Overall, the total number of posters we were able to find is relatively low, given that millions of scientific posters are estimated to be presented every year. Posters are mainly shared through institutional and general repositories, and are then scattered across several other repositories. We were able to count posters on less than half of the platforms retained for our analysis, and get the year-by-year count on less than 5% of them. This suggests that even when they are shared, the discoverability of posters (ability to filter, search, etc.) remains limited.

## FAIR practices

### Persistent identifiers

Of the 86 platforms retained for our analysis that contain posters, we found that 63 (73.2%) are issuing Digital Object Identifiers (DOIs) for posters. These DOI-issuing platforms account for 62,830 (41.8%) of the posters we counted. All other platforms seem to either assign a local identifier or not assign any identifier. This indicates that while DOI seems to be the de facto standard for persistent identification of posters, its adoption is inconsistent across repositories.

### Metadata collected by platforms

A list of key structured metadata collected by each platform is provided in **Table 1**. We found that Zenodo collects 33 unique structured metadata elements, of which 4 are mandatory. In particular, Zenodo has fields where the user can provide information about the conference where the poster was presented, including the name of the conference, dates, website link, and the poster's session. Figshare collects 16 unique structure metadata elements, of which 9 are mandatory. However, Figshare lacks fields for conference details.

### Completeness of general metadata provided by researchers

Overall, the completeness of general metadata varies substantially across metadata elements and between repositories, ranging from 100% for some to 0% for others (**Fig. 3**). Across both Zenodo (n = 24,734) and Figshare (n = 14,387), metadata elements that are mandatory at submission, such as title and author names, are consistently provided in all poster records. Among metadata elements that are optional, completeness varies widely. We observe that some optional metadata have high completeness on Zenodo. They seem to be ones that provide direct value to the authors, like their affiliations (79% completeness; 19,630 posters), or are standard when sharing research outcomes, like license (98%; 24,349) and description (96%; 23,787). In contrast, other optional metadata show consistently low completeness across both repositories, including author ORCID (26% on Zenodo and 15% on Figshare; 6,338 and 2,072 posters, respectively), funding information (19% and 27%; 4,803 and 3,984), and references (8% and 0%; 1,867 and 0). We observe the effect of mandatory requirements the most clearly when looking at metadata elements that are mandatory or auto-completed in one repository but optional in the other. For example, keywords and version number are mandatory and auto-completed, respectively, in Figshare and are present in nearly all Figshare records, whereas the same metadata are optional in Zenodo and show substantially lower completeness (54% and 8%, respectively; 13,401 and 1,874 posters). Together, these findings show that making metadata mandatory at submission leads to substantially higher completeness, and that repository design decisions likely play a central role in shaping the availability of metadata for posters.

**Completeness of poster-related metadata**

We then analyzed the completeness of poster-related metadata that authors can optionally provide when sharing a poster on Zenodo (not supported by Figshare). Our findings are summarized in **Fig. 4**. We observed that less than half of the posters have any conference-related metadata (48%; 11,765 of 24,734 posters). Some metadata are more commonly provided, such as conference name (47%; 11,648), conference dates (45%; 11,096), conference location (41%; 10,127), and conference website (37%; 9,241), while others are less available, like conference acronyms (27.3%; 6,743), and conference session information (13.7%; 3,391). All seven conference-related metadata fields supported by Zenodo are provided in less than 2% of the posters (394 posters; 1.6%). These patterns closely mirror our earlier observations for optional metadata: when metadata are not required at submission, they are frequently omitted, even when they convey information that is critical for the discovery, contextualization, and reuse of posters.

## Poster reuse

Views were the most common engagement metric (mean = 246.6, median = 95), followed by downloads (mean = 120.2, median = 62). Citations were rare, with a mean of 0.49 and a median of 0 (**Table 2)**. The proportion of posters with zero engagement was 11.5% for views, 11.7% for downloads, and 94.9% for citations. Among the 1,955 posters that received citations, the mean citation count was 9.5 with a median of 2. Engagement concentration was substantial across all metrics. The top 1% of posters accounted for 18.7% of views, 15.4% of downloads, and 86.3% of citations. The top 10% accounted for 53.2% of views, 46.9% of downloads, and 100% of citations (**Fig. 5**).

Bivariate analysis revealed significant positive associations between views and multiple metadata quality indicators. The strongest correlations were observed for free-text keyword count ($\rho = 0.264$, $p < 0.001$), ORCID percentage ($\rho = 0.210$, $p < 0.001$), and total references ($\rho = 0.140$, $p < 0.001$). Similar patterns were observed for downloads. For citation status, references with persistent identifiers showed the strongest correlation ($r = 0.070$, $p < 0.001$). Multivariate Negative Binomial regression with fixed dispersion parameter identified the presence of a description as the strongest predictor of views (IRR = 2.20, 95% CI: 1.77-2.73, $p < 0.001$), indicating that posters with descriptions received 120% more views on average. Each SD increase in ORCID percentage was associated with 26% higher views (IRR = 1.26, 95% CI: 1.22-1.30, $p < 0.001$). Free-text keywords also showed a significant positive effect (IRR = 1.19, 95% CI: 1.13-1.26, $p < 0.001$). License information was not a significant predictor of views in the multivariate model (IRR = 1.31, 95% CI: 0.70-2.45, $p = 0.40$) (**Table 3**).

For citations, a two-part hurdle, referred to as Part 1 and Part 2, model was employed (**Table 4**). In the logistic model for citation probability (Part 1; n = 38,147; pseudo-$R^2 = 0.033$), the presence of a description increased the odds of being cited by 239% (OR = 3.39, 95% CI: 1.91-6.04, $p < 0.001$). Free-text keyword count had the strongest continuous effect (OR = 1.37, 95% CI: 1.29-1.45, $p < 0.001$ per SD). Among the 1,955 cited posters, no metadata predictors showed significant effects on citation count in the multivariate model, with the exception of free-text keywords, which showed a negative association (Part 2, IRR = 0.66, 95% CI: 0.54-0.81, $p < 0.001$) and funding information (IRR = 0.64, 95% CI: 0.45-0.93, $p = 0.02$).

For keywords, this negative Part 2 association contrasts with the positive one in Part 1, where a higher free-text keyword count was associated with greater odds of a poster being cited at all (OR = 1.37 per SD), consistent with the positive associations observed for views and downloads. The two parts of the hurdle model answer different questions: Part 1 models the probability of receiving any citation, whereas Part 2 models citation counts among the 1,955 posters that were already cited. We interpret Part 2 estimate cautiously, since citation counts within this small and highly overdispersed subset are likely driven by the underlying research and author visibility rather than by metadata. Keywords are therefore best understood as an aid to being discovered and to crossing the threshold of being cited, rather than as a driver of how many citations a cited poster ultimately accrues.

Overall, these results indicate that poster reuse is mostly seen through views and downloads rather than formal citations. Moreover, reuse is highly unequal as a small fraction of posters account for a large share of reuse activities, in particular citations. Metadata quality showed a consistent and statistically significant association with reuse. In particular, the presence of description, keywords, and author ORCIDs was linked to a higher view count and download count as well as a higher odds of being cited. Once a poster was cited, metadata generally did not explain variation in citation counts, suggesting that metadata mainly support initial discoverability.

## Discussion

We found that poster sharing is very fragmented. Discoverability of posters is very limited, as we found it difficult to filter by posters and/or by date on most of these platforms, making even a simple task such as finding out the number of posters hosted by platforms very challenging. Where counting the number of posters was possible, we observed that poster sharing appears to be increasing over time, but overall volumes remain small relative to the number of posters presented annually, suggesting that most posters are either not shared consistently or are shared in ways that are difficult to find and quantify.

DOI seems to be the de facto persistent identifier for posters, with nearly three-quarters of poster-hosting platforms issuing DOIs. However, DOI adoption was uneven, and DOI-issuing platforms accounted for less than half of the posters we could count. This inconsistency has important downstream implications as a lack of persistent identifiers can constrain reliable citation, reduce interoperability with scholarly infrastructure, and limit the ability to aggregate and evaluate poster outputs at scale. The lack of identifiers or the use of local identifiers we observed on many platforms likely contributes to the broader discoverability and tracking challenges we experienced.

Our metadata analysis highlights how repository design choices shape the completeness of metadata. Zenodo supports a broader set of structured metadata fields, including poster-specific conference metadata, while Figshare supports a smaller number and, in particular, lacks dedicated conference fields. Across both repositories, mandatory fields were universally present, whereas optional fields showed wide variability. Optional fields that provide direct value to the authors, like their affiliations and license, and description, are the most commonly provided. In contrast, other optional metadata, including author ORCID, funding information, and references, show consistently low completeness across both repositories. These results underscore that poster FAIRness is not only a function of author behavior but also of platform design.

Analysis of poster-specific metadata on Zenodo further illustrates the gap between what metadata platforms support and what authors actually provide when it is optional. Fewer than half of the posters from Zenodo included any conference-related metadata. This is consequential because conference metadata, like conference name and dates, are central to interpreting novelty, audience, and context, and they enable discovery pathways that are otherwise unavailable (e.g., searching by conference or year). This completeness suggests that even when repositories provide support, conference metadata will remain incomplete unless expectations are clarified and key fields are incentivized or required.

Our analysis of poster reuse shows that reuse is mostly seen through views and downloads rather than formal citations. Most of the posters had zero citations, while views and downloads were more common but still unevenly distributed. A small fraction of posters accounted for a disproportionate share of engagement, especially citations. Importantly, metadata quality was consistently associated with higher engagement and with the probability of being cited. Descriptions, keywords, and ORCIDs were among the strongest predictors of reuse. This supports the role of metadata as a discoverability layer: richer, more complete poster records are easier to find, assess, and attribute, which translates into greater reuse.

Several limitations should be noted. First, despite our extensive search methods, we may not have identified all platforms where posters are shared. Second, platform-level poster count may be overestimated because we noticed that some records marked as “poster” did not include a poster file (some only included an abstract, for instance). However, this potential overestimate reinforces our central observation that posters are not consistently shared. Third, our structured metadata analysis was limited to Zenodo and Figshare due to their convenient API, which may not necessarily generalize to other platforms. Nevertheless, these two repositories are widely used and designed to support persistent identifiers and FAIR-aligned metadata workflows, making them a useful lens on current poster metadata practices. A further limitation is that our analysis is associational and cannot establish the mechanism by which richer metadata leads to greater discovery and reuse. We nonetheless offer plausible pathways. Descriptions and keywords likely help by giving both external search engines and a repository's own search more text to match against a user's query. Author ORCIDs may act through several routes at once: surfacing records on authors' ORCID and scholarly profiles, generating referral traffic and inbound links that can raise a record's ranking in search, and enabling reliable attribution and aggregation across systems. We cannot rule out reverse causality or confounding by author diligence, since an author who completes metadata may also be more likely to promote a poster. These pathways should therefore be read as hypotheses that motivate future controlled study rather than established causal channels.

Despite these limitations, our findings suggest several practical steps for all stakeholders to improve poster sharing and reuse and, ultimately, increase the scientific value of posters as research objects. First, research communities need to raise awareness that posters are legitimate scholarly outputs and promote norms for sharing them in trusted repositories. They should develop lightweight guidelines that specify where posters should be shared and define a minimal metadata package needed for discovery and attribution (e.g., abstract/description,

keywords, author identifiers such as ORCID, funding information, and conference information). Based on our learning from data sharing, we believe that tools that facilitate metadata extraction and sharing could help greatly in the adoption of such guidelines. Second, repositories need to translate these norms into practice. Since mandatory and auto-provided fields are completed far more often than optional ones, repositories should consider key metadata mandatory for posters. Repositories should streamline ORCID entry, for example via ORCID API lookups at submission, given its consistent link to higher views. Because posters are inherently tied to a conference, repositories should require conference names at minimum. Repositories should also prioritize DOI issuance. Third, funders, institutions, and researchers can accelerate adoption not just by encouraging sharing broadly, but also recognizing posters sharing in reporting and evaluation. We believe these steps can take poster from being the neglected child of scientific research to being a first class output that enhances the pace of discoveries and collaborations.

## Data availability

The data associated with this manuscript consists of several Excel files and JSON/NDJSON files that contain details about the platform search, metadata analysis, and poster reuse. Since no FAIR guidelines were found for structuring such data, we structured it according to the SPARC Data Structure (SDS), which provides a broad data and metadata structure to organize biomedical research data according to the FAIR principles.[34] The SPARC data curation software SODA for SPARC was used to organize the data and prepare the metadata files.[35,36] The dataset is archived on Zenodo.[37] This data is shared under the permissible Creative Commons Attribution 4.0 International (CC-BY) license.

## Code availability

The code associated with this manuscript consists of multiple Jupyter notebooks (mentioned in the Methods), which contain the code used to find platforms, retrieve metadata from Zenodo and Figshare, and analyze our data. These notebooks are maintained in a GitHub repository called “poster-sharing-reuse-paper-code” in the “fairdataihub” GitHub organization. The notebook was made FAIR according to the FAIR-BioRS guidelines using Codefair. Accordingly, the version associated with this manuscript (v1.0.0) is archived on Zenodo.[38] The notebook and associated files are shared under the permissible MIT license.

## Author Contributions

All authors contributed to data curation, table preparation, manuscript editing, and writing. B.P. and A.G. developed the script files, and J.O.N. contributed substantial modifications and improvements. P.M.K. and B.P. prepared the initial manuscript and spreadsheet file.

## Competing Interests

The author(s) declare no competing interests.

## Funding

This work was supported by the The Navigation Fund through grant 10.71707/rk36-9x79[39] and the US National Institutes of Health (NIH) through grant OT2OD032644.

## Figures

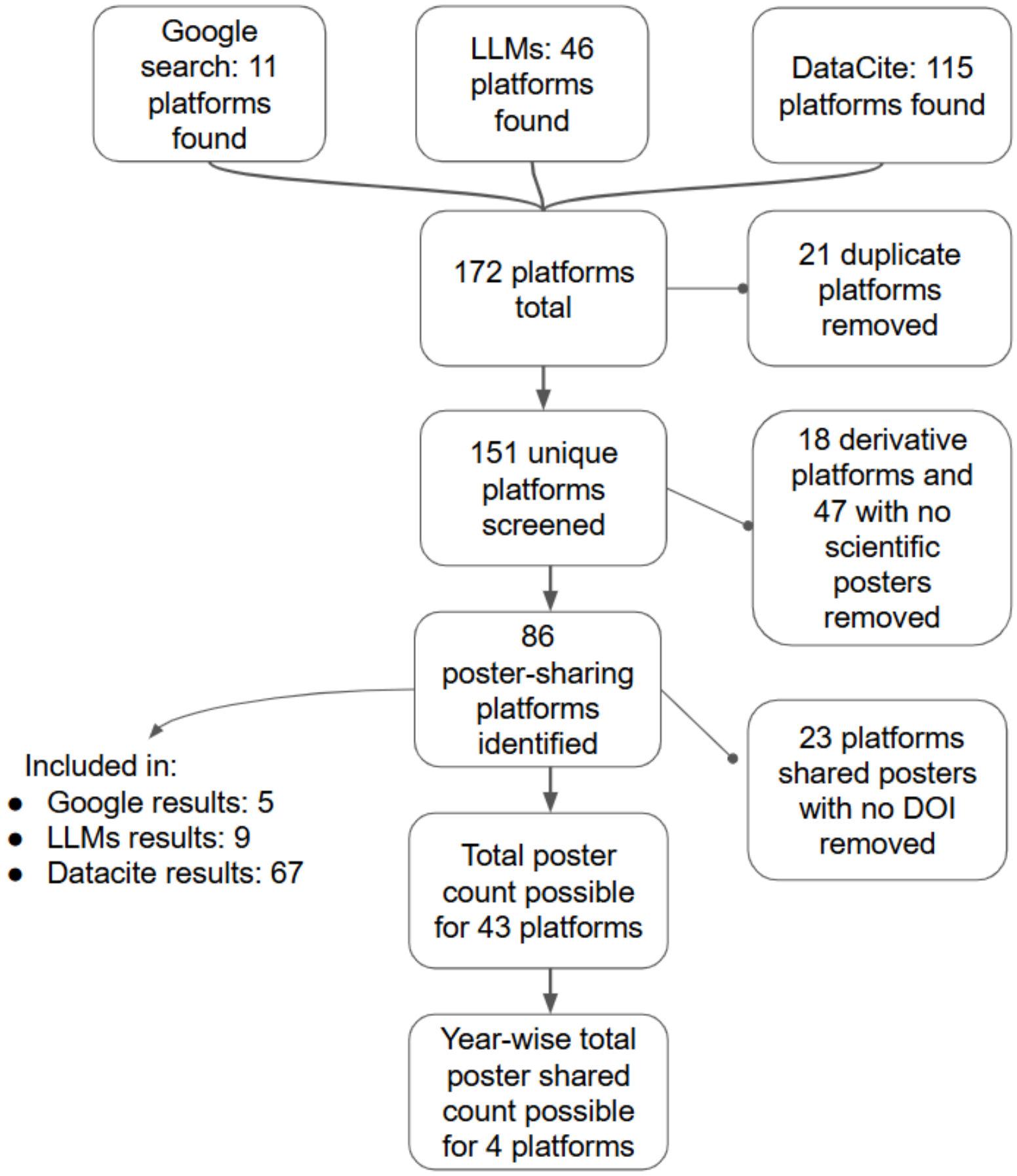

**Figure 1**. Overview of our process to identify platforms where posters are shared.

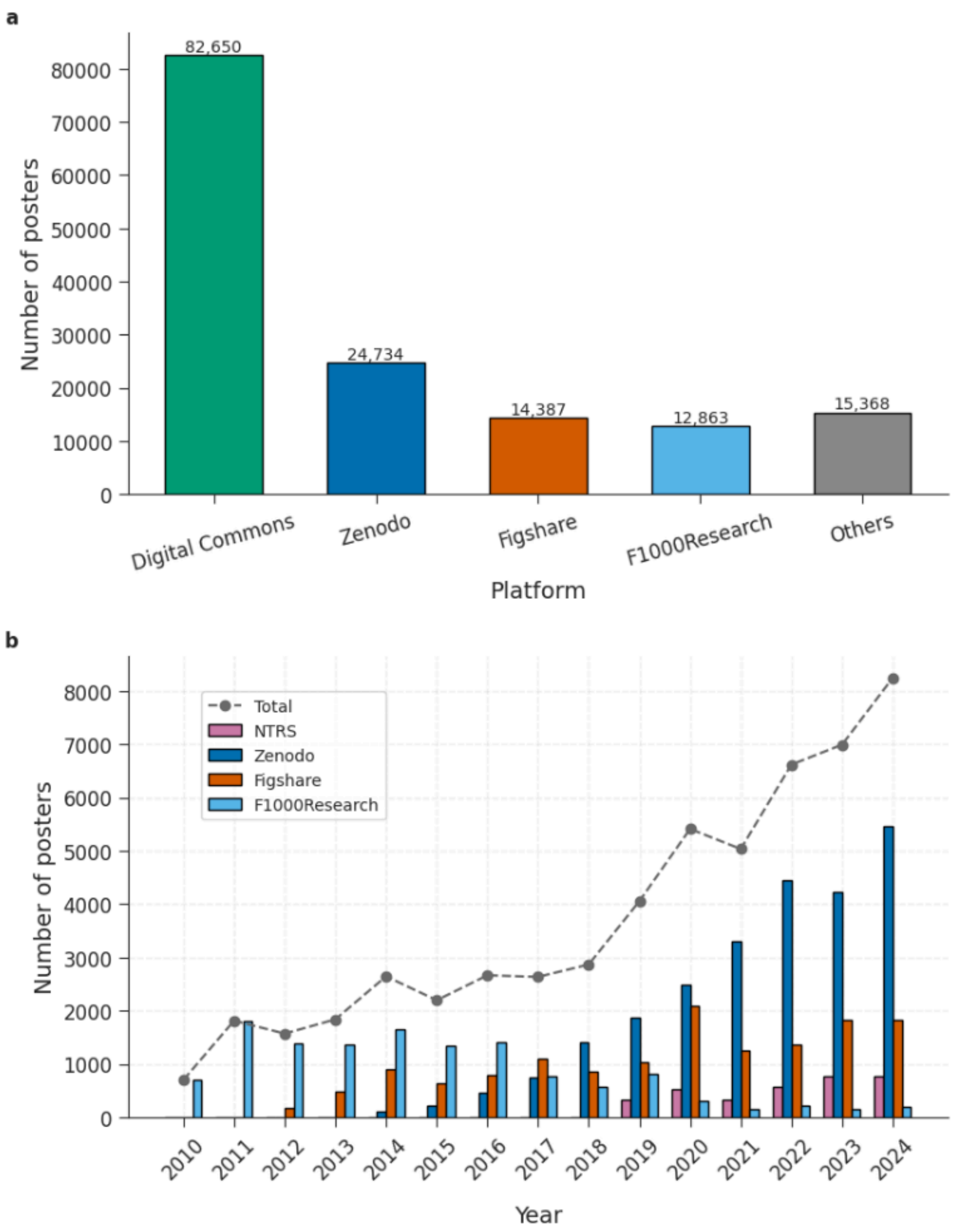

**Figure 2**. Poster count across platforms. (a) Number of posters across platforms with the highest count. (b) Evolution of the number of posters shared across the 4 platforms, where a year-by-year count was possible.

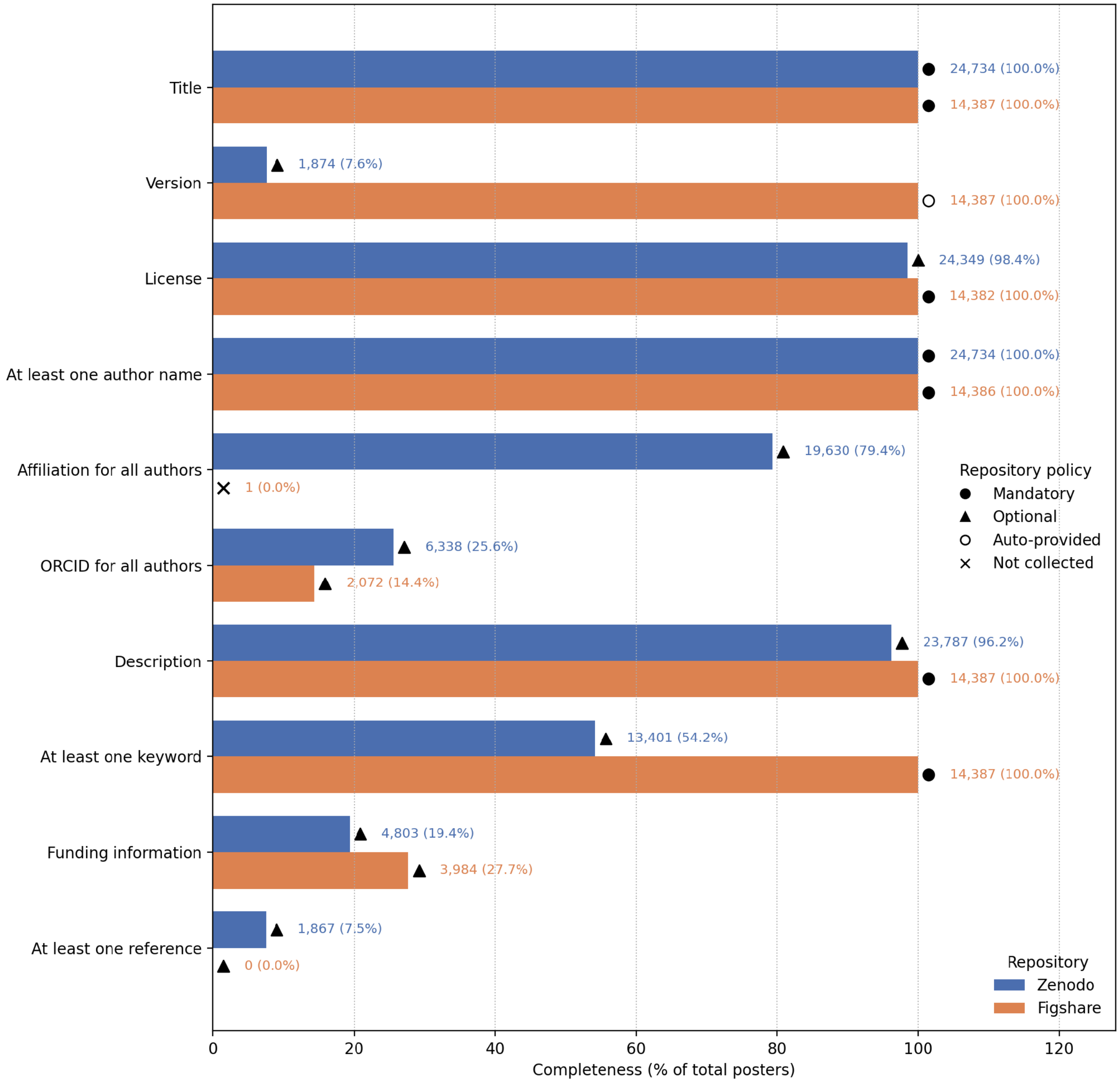

**Figure 3**. Completeness of general metadata provided by researchers sharing posters on Zenodo and Figshare (n = 24,734 and 14,387, respectively). Bars are labeled with the number of posters and the corresponding percentage of the total for each field.

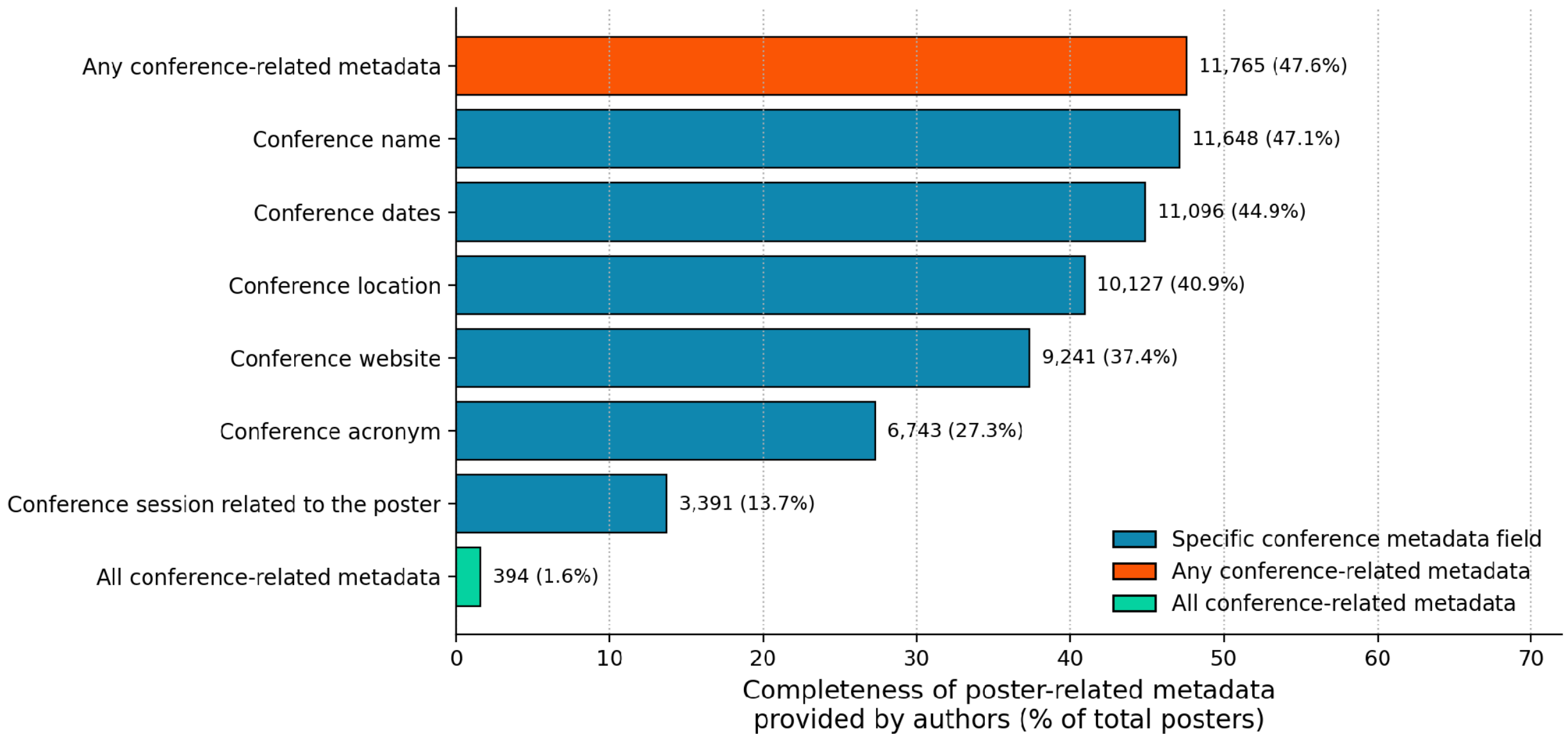

**Figure 4**. Completeness of poster-related metadata provided by researchers sharing posters on Zenodo (n = 24,734).

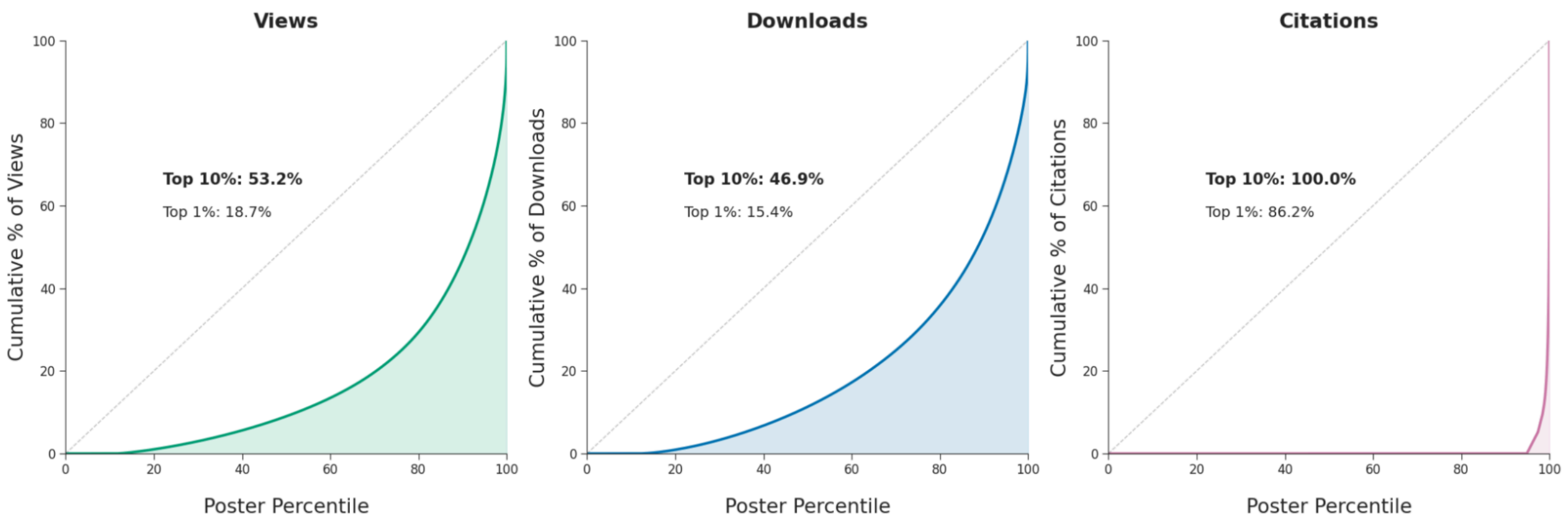

**Figure 5**. Cumulative distribution of engagement metrics. Each panel shows the cumulative percentage of total engagement (views, downloads, citations) captured by poster percentiles, sorted from lowest to highest engagement. The dashed diagonal represents perfect equality. The extreme concentration of citations (top 10% = 100%) contrasts with the more distributed views (top 10% = 53.2%) and downloads (top 10% = 46.9%).

## Tables

**Table 1**. Comparison of structured metadata fields captured by Zenodo and Figshare. Asterisks (*) indicate mandatory fields.

| Key structured metadata | Zenodo | Figshare |
|---|---|---|
| Title of the record | yes* | yes* |
| Version of the record | yes | no |
| Description/abstract of the record | yes | yes* |
| Other descriptions | yes | no |
| Name of license | yes | yes* |
| Full name of author | yes* | yes* |
| Affiliation of author | yes | no |
| Unique identifier of the author | yes | yes |
| Keyword as free text | yes | yes* |
| Keywords from controlled vocabularies | yes | yes* |
| Funding/grant number | yes | no |
| Title of the funding/grant | yes | yes* |
| URL with more information about the funding/grant | yes | no |
| Funder name | yes | no |
| Title of the conference where the record was presented | yes | no |
| Website of the conference | yes | no |
| Dates of the conference | yes | no |
| Related items for items where the identifier is not known | yes | no |
| Identifier of related items with a known identifier | yes | yes |
| Relation to the record of related items with a known identifier | yes | yes |
| List of files in the record | yes* | yes* |

**Table 2**. Statistics characterizing the distributions of views, downloads, and citations.

| ***Metric*** | ***N*** | ***Mean*** | ***Median*** | ***Std Dev*** | ***% Zero*** | ***Top 1%*** | ***Top 10%*** |
|---|---|---|---|---|---|---|---|
| Views | 38,147 | 246.6 | 95 | 1,135.2 | 11.5% | 18.7% | 53.2% |
| Downloads | 38,147 | 120.2 | 62 | 585.9 | 11.7% | 15.4% | 46.9% |
| Citations | 38,147 | 0.49 | 0 | 29.9 | 94.9% | 86.3% | 100.0% |

**Table 3**. Multivariate Predictors of Poster Views. Negative Binomial regression with fixed dispersion parameter (α = 10.0). Continuous predictors are standardized prior to modeling. Results reported as Incidence Rate Ratios (IRR) with 95% confidence intervals. *** p < 0.001, ** p < 0.01, * p < 0.05.

| Predictor | IRR | 95% CI | p-value |
|---|---|---|---|
| Has description | 2.20 | [1.77, 2.73] | <0.001*** |
| Has license | 1.31 | [0.70, 2.45] | 0.40 |
| ORCID % (per SD) | 1.26 | [1.22, 1.30] | <0.001*** |
| Affiliation % (per SD) | 0.86 | [0.83, 0.90] | <0.001*** |
| Description words (per SD) | 1.04 | [1.01, 1.08] | 0.010** |
| Free-text keywords (per SD) | 1.19 | [1.13, 1.26] | <0.001*** |
| Has conference acronym | 1.08 | [0.96, 1.23] | 0.20 |
| Has conference place | 1.21 | [1.00, 1.47] | 0.05* |
| Has conference title | 0.61 | [0.46, 0.81] | <0.001*** |
| Has conference session | 0.73 | [0.63, 0.85] | <0.001*** |
| Has funding info | 0.90 | [0.84, 0.98] | 0.009** |

*Negative Binomial regression with fixed dispersion parameter (α estimated from intercept-only model). IRR = Incidence Rate Ratio. N = 38,147. *p<0.05, **p<0.01, ***p<0.001.*

**Table 4.** Two-Part Hurdle Model for Citations. Part 1: Logistic regression predicting probability of any citation (n = 38,147). Results reported as Odds Ratios (OR). Part 2: Negative Binomial regression with fixed dispersion parameter predicting citation count among cited posters (n = 1,955). Results reported as Incidence Rate Ratios (IRR). *** p < 0.001, ** p < 0.01, * p < 0.05.

| **Predictor** | **Part 1: OR** | **p** | **Part 2: IRR** | **p** |
|---|---|---|---|---|
| Has description | 3.39 | <.001*** | 3.41 | 0.19 |
| Has license | 5.32 | 0.10 | 6.54 | 0.57 |
| Free-text keywords (per SD) | 1.37 | <.001*** | 0.66 | <.001*** |
| Refs with identifiers (per SD) | 1.14 | <.001*** | 0.96 | 0.47 |
| Has conference acronym | 1.38 | <.001*** | 0.78 | 0.34 |
| Has conference dates | 2.23 | 0.00** | 1.34 | 0.73 |
| Has funding info | 0.80 | <.001*** | 0.64 | 0.02* |
| Has conference session | 0.68 | <.001*** | 0.90 | 0.75 |

*Part 1: Logistic regression for P(cited), N = 38,147. Part 2: Negative Binomial with fixed dispersion for count among cited, N = 1,955. SD = Standard Deviation, OR = Odds Ratio, IRR = Incidence Rate Ratio. *p<0.05, **p<0.01, ***p<0.001.*

**Table 5**. Metadata variables retained for the analysis of correlation between metadata, views, downloads, and citations.

| Variable | Type | Description |
|---|---|---|
| authors_affiliation_percentage | Float | Percentage of authors with institutional affiliations |
| authors_orcid_percentage | Float | Percentage of authors with ORCID identifiers |
| description_words_count | Integer | Number of words in the description |
| keywords_freetext_count | Integer | Number of free-text keywords |
| keywords_controlled_vocabularies_count | Integer | Number of controlled vocabulary keywords |
| total_keywords_count | Integer | Sum of all keywords |
| references_with_identifiers_count | Integer | References with DOIs/identifiers |
| references_no_identifiers_count | Integer | References without identifiers |
| total_references_count | Integer | Total references |
| has_description | Boolean | Poster has a description |
| has_license | Boolean | License provided |
| has_funding_info | Boolean | Funding information provided |
| has_conference_acronym | Boolean | Conference acronym present |
| has_conference_dates | Boolean | Conference dates provided |
| has_conference_place | Boolean | Conference location provided |
| has_conference_session | Boolean | Conference session provided |
| has_conference_title | Boolean | Conference title provided |
| has_conference_website | Boolean | Conference website provided |